\documentclass[12pt]{article}

\pdfoutput=1
\usepackage{graphicx} 
\usepackage{xspace}
\usepackage{url}
\usepackage{comment}
\usepackage{amsmath}
\usepackage{amssymb}
\usepackage{fancyhdr}  
\usepackage{hepunits}
\usepackage[usenames,dvipsnames]{color}  
\usepackage{enumitem}
\usepackage{multirow}
\usepackage{lineno}
\usepackage{pdfpages}

\usepackage[a4paper, total={17cm, 22cm}, headheight=0.6cm]{geometry}
\usepackage{subcaption}  
\usepackage[pdftex,bookmarks,hidelinks]{hyperref}

\usepackage[normalem]{ulem}
\newcommand{\rms}{RMS\xspace} 

\def\argon40{${}^{40}$Ar}       
\def\Ar39{$^{39}$Ar}
\def\Cl40{$^{40}$Cl}
\def\K40{$^{40}$K}
\def\B8{$^{8}$B}
\newcommand{\lsim}{{\;\raise0.3ex\hbox{$<$\kern-0.75em\raise-1.1ex\hbox{$\sim$}}\;}}
\newcommand{\gsim}{{\;\raise0.3ex\hbox{$>$\kern-0.75em\raise-1.1ex\hbox{$\sim$}}\;}}
\newcommand{\beq}{\begin{equation}}
\newcommand{\eeq}{\end{equation}}
\newcommand{\bea}{\begin{eqnarray}}
\newcommand{\eea}{\end{eqnarray}}

\mathchardef\minus="002D

\DeclareSIUnit \c {$c$}
\DeclareSIUnit\magn{$\times$}
\DeclareSIUnit\min{min}
\DeclareSIUnit\hr{hr}
\DeclareSIUnit\hrs{hrs}
\DeclareSIUnit\week{week}
\DeclareSIUnit\month{mo}
\DeclareSIUnit\months{mos}
\DeclareSIUnit\year{yr}
\DeclareSIUnit\years{years}
\DeclareSIUnit\yr{yr}
\DeclareSIUnit\standard{std}
\DeclareSIUnit\str{sr}
\DeclareSIUnit\ppm{ppm}
\DeclareSIUnit\ppb{ppb}
\DeclareSIUnit\ppt{ppt}
\DeclareSIUnit\pe{PE}
\DeclareSIUnit\spe{SPE}
\DeclareSIUnit\pdm{PDM}
\DeclareSIUnit\ev{events}
\DeclareSIUnit\ct{counts}
\DeclareSIUnit\neutron{\mbox{$n$}}
\DeclareSIUnit\smp{samples}
\DeclareSIUnit\Sample{S}
\DeclareSIUnit\ch{ch}
\DeclareSIUnit\hit{hit}
\DeclareSIUnit\hits{hits}
\DeclareSIUnit\bin{(\mbox{5-PE}~bin)}
\DeclareSIUnit\sgm{\mbox{$\sigma$}}
\DeclareSIUnit\rms{RMS}
\DeclareSIUnit\keVee{\mbox{keV$_{e{\rm e}}$}}
\DeclareSIUnit\keVr{\mbox{keV$_{\rm nr}$}}
\DeclareSIUnit\eVee{\mbox{eV$_{\rm ee}$}}
\DeclareSIUnit\eVr{\mbox{eV$_{\rm nr}$}}
\DeclareSIUnit\ph{photon}
\DeclareSIUnit\el{\mbox{$e^-$}}
\DeclareSIUnit\pm{\mbox{PMT}}
\DeclareSIUnit\pixel{\mbox{pixel}}
\DeclareSIUnit\inch{''}
\DeclareSIUnit\foot{'}
\DeclareSIUnit\bit{bit}
\DeclareSIUnit\sample{samples}
\DeclareSIUnit\barn{barn}
\DeclareSIUnit\bara{bar}
\DeclareSIUnit\bar{bar}
\DeclareSIUnit\barg{barg}
\DeclareSIUnit\mlardepth{\mbox(meter~of~\LAr~depth)}
\DeclareSIUnit\Curie{Ci}
\DeclareSIUnit\PSI{psi}
\DeclareSIUnit\psia{psia}
\DeclareSIUnit\atm{atm}
\DeclareSIUnit\psf{psf}
\DeclareSIUnit\pcf{pcf}
\DeclareSIUnit\parsec{pc}
\DeclareSIUnit\cps{cps}
\DeclareSIUnit\slpm{\SI{}{\liter\per\minute}}
\DeclareSIUnit\rpm{rpm}
\DeclareSIUnit\mwe{\mbox{m.w.e.}}
\DeclareSIUnit\liveday{\mbox{live-days}}
\DeclareSIUnit\days{\mbox{days}}
\DeclareSIUnit\miles{\mbox{miles}}
\DeclareSIUnit\lumens{\mbox{lm}}
\DeclareSIUnit\degreeC{\mbox{$^{\circ}$C}}
\DeclareSIUnit\degreeF{\mbox{$^{\circ}$F}}
\DeclareSIUnit\electron{\mbox{$e^-$}}
\DeclareSIUnit\Euro{\mbox{\euro}}
\DeclareSIUnit\cph{cph}
\DeclareSIUnit\neq{neq}
\DeclareSIUnit\normal{\mbox{N}}
\DeclareSIUnit\USD{\mbox{\$}}
\DeclareSIUnit\Vpercm{\mbox{V/cm}}
\DeclareSIUnit\kV{\mbox{kV}}

\DeclareSIUnit \mm {\milli\meter}
\DeclareSIUnit \cm {\centi\meter}
\DeclareSIUnit \us {\micro\second}
\DeclareSIUnit \ms {\milli\second}
\DeclareSIUnit \pA {\pico\ampere}
\DeclareSIUnit \pC {\pico\coulomb}
\DeclareSIUnit \fC {\femto\coulomb}
\DeclareSIUnit \fF {\femto\farrad}
\DeclareSIUnit \pF {\pico\farrad}
\DeclareSIUnit \mV {\milli\volt}
\DeclareSIUnit \kV {\kilo\volt}
\DeclareSIUnit \V {\volt}
\DeclareSIUnit \GOhm {\giga\ohm}
\DeclareSIUnit \MOhm {\mega\ohm}
\DeclareSIUnit \ton {\tonne}
\DeclareSIUnit \kton {\kilo\tonne}
\DeclareSIUnit \kt {\kilo\tonne}
\DeclareSIUnit \Mt {\mega\tonne}
\DeclareSIUnit \eV {\electronvolt}
\DeclareSIUnit \keV {\kilo\electronvolt}
\DeclareSIUnit \MeV {\mega\electronvolt}
\DeclareSIUnit \GeV {\giga\electronvolt}
\DeclareSIUnit \km {\kilo\meter}
\DeclareSIUnit \kW {\kilo\watt}
\DeclareSIUnit \MW {\mega\watt}
\DeclareSIUnit \MHz {\mega\hertz}
\DeclareSIUnit \kHz {\kilo\hertz}
\DeclareSIUnit \mrad {\milli\radian}
\DeclareSIUnit \year {year}
\DeclareSIUnit \POT {POT}
\DeclareSIUnit \sig {$\sigma$}
\DeclareSIUnit\parsec{pc}
\DeclareSIUnit\lightyear{ly}
\DeclareSIUnit\foot{ft}
\DeclareSIUnit\ft{ft}

\title{%
DUNE Software and Computing Research and Development  \\ \bigskip
\large Input to the European Strategy for Particle Physics - 2026 Update }

\author{The DUNE Collaboration
\footnote{Contact persons: Sergio Bertolucci (Sergio.Bertolucci@cern.ch), Sowjanya Gollapinni (sowjanya@lanl.gov)}
}

\date{\today}

\begin{document}

\maketitle

\begin{abstract}
The international collaboration designing and constructing the Deep Underground Neutrino Experiment (DUNE) at the Long-Baseline Neutrino Facility (LBNF) has developed a two-phase strategy toward the implementation of this leading-edge, large-scale science project. The ambitious physics program of Phase~I and Phase~II of DUNE is dependent upon deployment and utilization of significant computing resources, and successful research and development of software (both infrastructure and algorithmic) in order to achieve these scientific goals. This submission discusses the computing resources projections, infrastructure support, and software development needed for DUNE during the coming decades as an input to the European Strategy for Particle Physics Update for 2026.

The DUNE collaboration is submitting four main contributions to the 2026 Update of the European Strategy for Particle Physics process. This submission to the “Computing” stream focuses on DUNE software and computing. Additional inputs related to the DUNE science program, DUNE detector technologies and R\&D, and European contributions to Fermilab accelerator upgrades and facilities for the DUNE experiment, are also being submitted to other streams.
\end{abstract}

\thispagestyle{empty}

\newpage
\pagenumbering{arabic}

\section{Introduction}

  The overriding physics goals of the DUNE experiment are the search for leptonic CP violation, the search for nucleon decay as a signature of a Grand Unified Theory underlying the Standard Model, the observation of supernova burst neutrinos from supernovae in our galaxy, and studies of solar neutrinos. DUNE will consist of modular Far Detectors (FDs) located about \SI{1.5}{km} underground at Sanford Underground Research Facility (SURF), \SI{1300}{\km} from Fermi National Accelerator Facility (FNAL), and a Near Detector (ND) located on site at FNAL in Illinois. The DUNE detectors will be exposed to the world's most intense neutrino beam originating at LBNF. The FDs will be sensitive to neutrinos from numerous sources, with the detector design and location optimized for detection of neutrinos from LBNF. The high-precision near detector, \SI{574}{m} from the neutrino source on the FNAL site, will be used to characterize the intensity and energy spectrum of this wideband beam along with studying neutrino-nucleon interactions among other physics goals. DUNE also has an active program of prototype operations (e.g. ProtoDUNE at CERN) that will be produce significant data prior to the start of Far Detector operations near the end of this decade. 
  
\subsection{DUNE's Computing Challenges}

The DUNE physics program drives several detector characteristics that pose novel computing challenges.  While the overall data volumes are smaller than those routinely handled by the LHC experiments, the large, open far detector design and broad physics goals present computing challenges different from those addressed by LHC experiments. DUNE offline computing faces six major challenges that must be addressed in the near future, some of which are unique to DUNE and others shared widely by HEP experiments.

\begin{itemize}
\item{\bf Large event records and memory footprints ---}  DUNE Far Detector (FD) events present formidable challenges for reconstruction on typical HEP processing systems, with the data volume of a beam-synchronous trigger record from a single FD module potentially as large as 8 GB. Additionally, extended-time readout of FD modules ($\approx$100 seconds of continuous readout and the corresponding $\approx$150 TB of data) for Supernova neutrino observations presents unique challenges for both the computing infrastructure and the production processing framework. Along with data reconstruction, the simulation of such a large, open detector creates significant challenges for traditional HTC computing architectures. Efficient processing of DUNE data will require careful attention to data formats and substantial redesign of the processing framework to allow sequential processing of chunks of data that are correlated in both time and space.

\item{\bf Storing and processing data on heterogeneous international, cloud, and HPC resources ---} DUNE depends on the combined resources of the collaboration for large-scale storage and processing of data.   Tools for using shared resources ranging from small-scale clusters to dedicated HPC systems need to be developed and maintained.   Fortunately, HEP, through the WLCG, OSG, and HSF, has a well-developed ecosystem of tools that allow reasonably transparent use of collaboration computing resources, but the development of features must be optimized for the flatter computing model utilized by DUNE.

\item{\bf Maintaining operational efficiency ---}  
There is a large suite of activities, that,   
while not necessarily novel, 
still needs to be done over the full lifetime of the experiment.  These activities include database design and operations, security updates, code management, documentation, training, and user support.  One example is the continuing evolution of operating systems and security requirements.  These require constant modifications to working systems to maintain operations across the lifetime of the experiment, which will be greater than 20 years. Engagement and support from computing expertise around the world will be necessary in order to maintain a stable and resilient global computing infrastructure.

\item{
\bf Machine Learning ---} Use of machine learning techniques can greatly improve simulation, reconstruction and analysis of data. However, integration of ML techniques into a software ecosystem of the size and complexity of a large HEP experiment requires substantial effort beyond the original demonstration.  Access to considerable resources for ML training, optimized data format, and algorithmic reproducibility on varied architectures are all considerable concerns for DUNE as with other HEP experiments. Adapting to various architectures (both accelerators and facility specific) will be a challenge that needs to be addressed and hopefully can come from common solutions.

\item{\bf Integration of HPC resources into Computing Models ---} 
Much of DUNE's raw data is generated by large numbers of almost-identical detector components with reasonably uniform data structures.  This makes our data well-suited for simulation and reconstruction using co-processors. Improved projections of needs and negotiation access to large amounts of HPC resources, especially if they are used for near real-time data processing, will take considerable effort. Additionally, as these resources become a larger and larger part of resource needs, the ability to change from annual allocations to longer term access agreements will be critical to planning. 

\item{\bf Efficient and sustainable use of resources ---} Due to the large data volume from the FD and complexity of the ND interactions, DUNE will use substantial computing resources.  Historically, the main concern has been financial - getting the most computing possible within a given budget. However, our activities also have environmental impact, through energy consumption and the creation, use, and disposal of hardware. To make our efforts more sustainable, a driving consideration in the design of our systems is efficient use of storage, CPU, ARM, GPU, and other resources.  This includes not just optimization of our major workflows, but also documentation and training of end-users in efficient and error-resistant practices to avoid needless reprocessing.

\end{itemize}

Additional details about these issues, and additional planning for DUNE Software and Computing research and development, are discussed throughout the DUNE Software and Computing Conceptual Design Report\cite{offline-cdr}.

\section{DUNE Computing Resources Projections}

\begin{figure}
    \centering
    \includegraphics[width=0.5\linewidth]{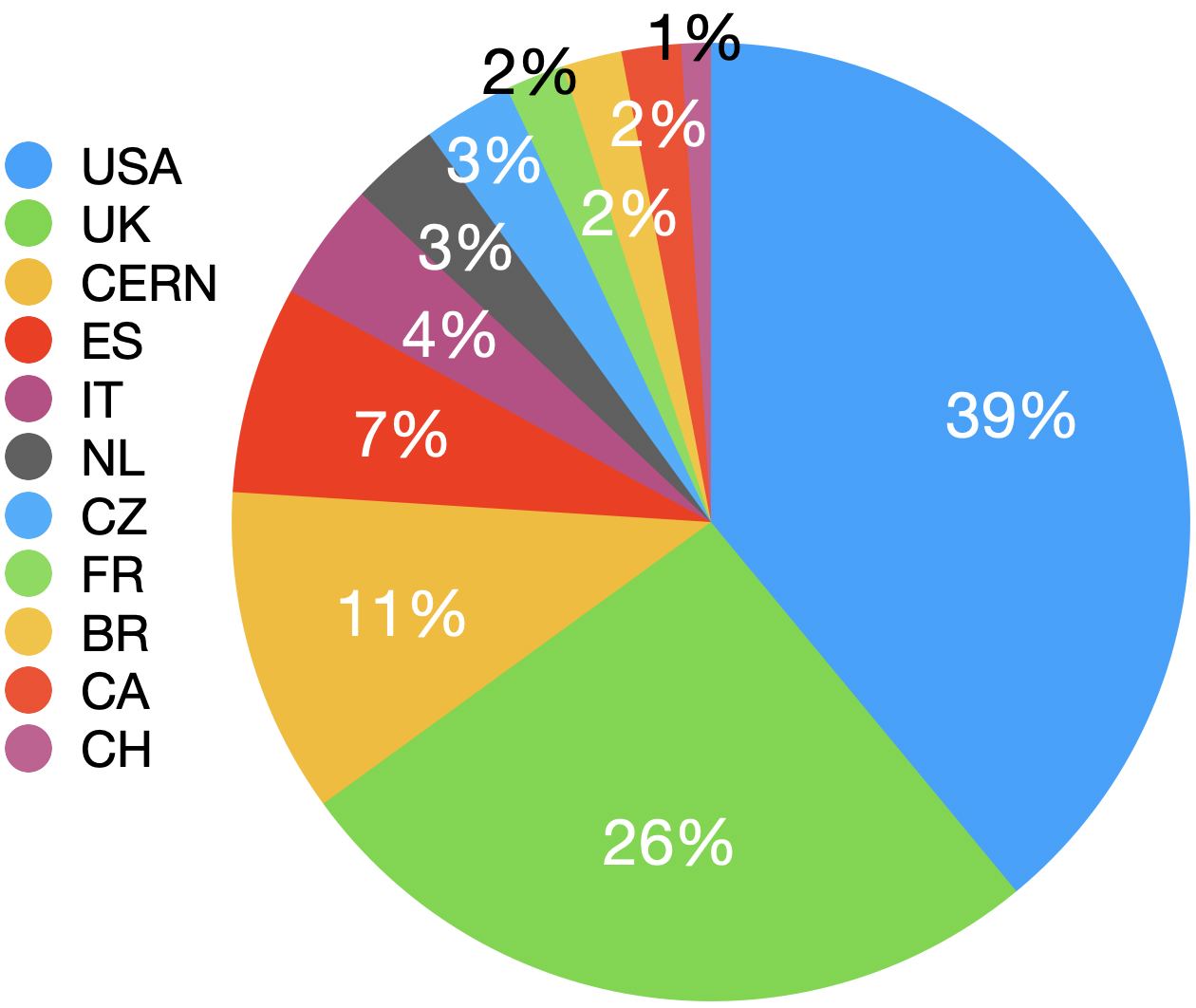}
    \caption{The fraction of CPU resources utilized in each nation for DUNE production processing campaigns for during 2024. These campaigns include reconstruction and simulation for DUNE Far Detectors and ProtoDUNE. The largest single national contribution came from the US, but the total European contribution was greater than 57\% of CPU processing.}
    \label{fig:prodsites}
\end{figure}

As an international collaboration,  contributions to the DUNE compute resources are provided by institutions around the globe. DUNE has chosen a flatten computing model compared with the tiered structure of the WLCG. At the same time, the long-term strategy for computing resources such that primary raw data storage to reside on tape at the large host labs (FNAL, CERN, and other national class facilities). For processing and storage of derived datasets, computing resources such as CPU and disk storage are contributed across both national facilities and collaborating
institutions. Current CPU contributions are provided from many sites worldwide, as shown in Figure~\ref{fig:prodsites}, while disk storage is currently concentrated at larger sites concentrated in the United States and Europe. This distributed collection of compute facilities provides efficient access to data for processing and a more flexible infrastructure through limited reliance on any one facility.

\begin{figure}
    \centering
    \includegraphics[width=0.7\linewidth]{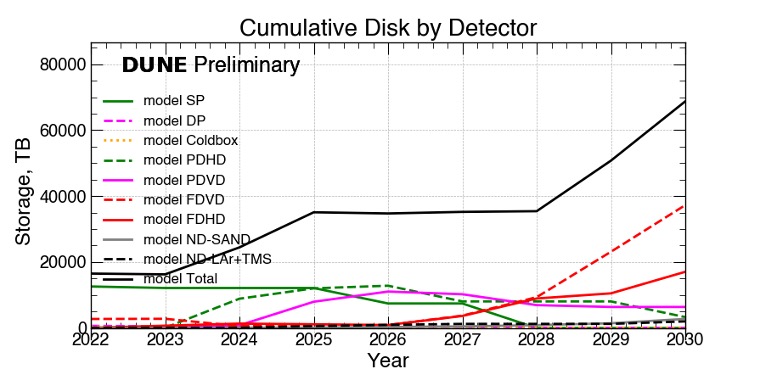}
    \caption{The DUNE estimates for disk storage needs through 2030 based upon the projections of data volumes from the ProtoDUNE Horizontal Drift (PDHD), ProtoDUNE Vertical Drift (PDVD), Far Detector Horizontal Drift (FDHD), Far Detector Vertical Drift (FDVD), Near Detector SAND (ND-SAND), and Near Detector Liquid Argon and Muon Spectrometer (ND-LAr+TMS).}
    \label{fig:disk_projection}
\end{figure}

\begin{figure}
    \centering
    \includegraphics[width=0.7\linewidth]{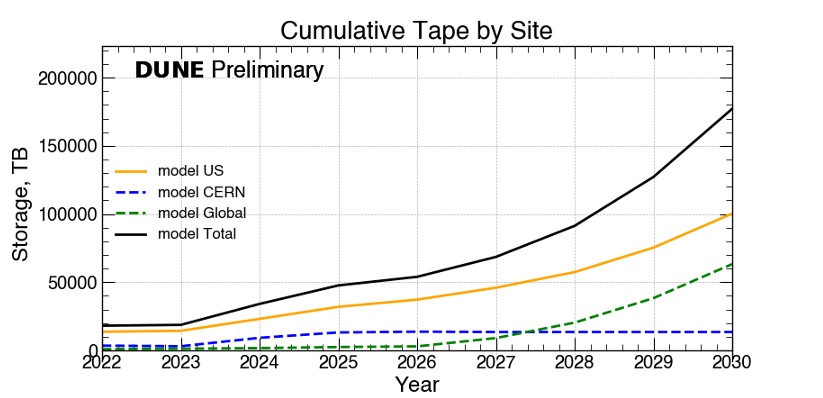}
    \caption{The DUNE estimates for tape storage needs through 2030 based upon the projections of data volumes from the ProtoDUNE Horizontal Drift (PDHD), ProtoDUNE Vertical Drift (PDVD), Far Detector Horizontal Drift (FDHD), Far Detector Vertical Drift (FDVD), Near Detector SAND (ND-SAND), and Near Detector Liquid Argon and Muon Spectrometer (ND-LAr+TMS).}
    \label{fig:tape_projection}
\end{figure}

\begin{figure}
    \centering
    \includegraphics[width=0.7\linewidth]{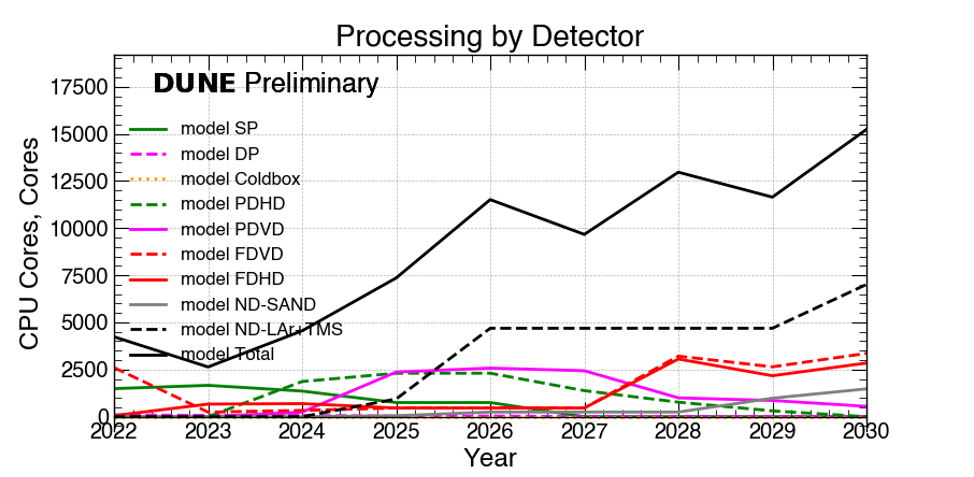}
    \caption{The DUNE estimates for CPU needs through 2030 based upon the projections of data volumes from the ProtoDUNE Horizontal Drift (PDHD), ProtoDUNE Vertical Drift (PDVD), Far Detector Horizontal Drift (FDHD), Far Detector Vertical Drift (FDVD), Near Detector SAND (ND-SAND), and Near Detector Liquid Argon and Muon Spectrometer (ND-LAr+TMS).}
    \label{fig:cpu_projection}
\end{figure}

To track and coordinate compute resources, DUNE has established an internal resource request process that updates resource projects yearly, and requests and tracks pledges from national facilities and collaborating institutions for storage and CPU compute. The actual need will change over time, especially due to changes in experimental operation, but our current experience is that compute resources needs have been successfully met with strong contributions from European partners, and especially infrastructure provided by CERN that are central to ProtoDUNE operations and data collection. These contributions are important to meeting DUNE's needs, especially concerning disk storage. The projections calculated in January 2025 for disk storage, tape storage, and CPU compute are shown, respectively, in Figures \ref{fig:disk_projection}, \ref{fig:tape_projection}, and \ref{fig:cpu_projection}. The support of CERN and European computing facilities and infrastructure is seen as critical to the success of DUNE's computing operations, and in turn, DUNE's physics goals.

\subsection{Bulk compute and storage using the Worldwide LHC Computing Grid}

DUNE has established an effective way of working using the 
Worldwide LHC Computing Grid (WLCG) infrastructure to access
compute and storage at participating institutes, and expects to
continue doing so through the remainder of ProtoDUNE data taking
and during physics data taking with the DUNE far and near detectors.

We envisage that bulk High Throughput Computing (HTC) grid 
resources of the type provided by WLCG will continue to be our 
primary computing solution to address our resource requirements, 
alongside the HPC resources described below. We also believe that utilization of these resources through WLCG software and infrastructure provides a long-term sustainability that is more robust than development of custom solutions for HTC computing.

Our strategy has been to reuse computing systems developed for the
LHC experiments where possible, to extend them to meet DUNE's unique
requirements where appropriate, and to develop new systems from
scratch where necessary. In particular, we have reused the glideInWMS
workload management system developed by Fermilab and adopted by CMS;
have adopted the Rucio data management system originally developed
by ATLAS and contributed DUNE staff effort to add core features 
needed by DUNE and others, and developed DUNE-specific modules
where necessary; we have worked with Fermilab's development of the
MetaCat file metadata catalogue which was created to address
requirements raised by DUNE and to work with Rucio; and we have
developed the justIN workflow management system to tie together
glideInWMS, Rucio, MetaCat and the neutrino applications which 
expect the just-in-time provisioning and data-access paradigm. 

This approach is working well to achieve the greatest impact in
facilitating the work of the collaboration with the limited staff
resources at hand, whilst presenting as much compatibility as
possible with the LHC experiments to site managers at institutes. 

We intend to continue this approach into the physics data taking
period, adjusting the specific choices of technology as new 
opportunities with processors, storage systems, and site 
architectures emerge.

We feel that DUNE provides useful diversity of timescales and 
approach, using significant computing resources across multiple 
continents, but with Fermilab rather than CERN as the host 
laboratory. For example, both DUNE and the LHC experiments are
planning to use WLCG tokens for authentication, but we need to 
adopt them on a shorter timescale due to Fermilab's constraints,
and this is providing useful experience before tokens become
essential to the running LHC experiments. 

\subsection{Accessibility of High Performance Computing Facilities}
As leadership class HPC facilities become a larger and larger part of the distributed computing model of HEP experiments, it is important for DUNE to take advantage of these facilities, as DUNE workflows have a number of computing tasks which have been shown to be good use cases for GPU- and accelerator-based computing. 
These workflows include, but are not limited to, processing data from the FDs, pattern recognition, particle identification, and final model parameter fits. Early studies have shown good results from a mixed CPU-GPU analysis~\cite{sonic}, which is a natural fit for some of the current and next generation of HPC facilities across Europe\cite{eu_hpc}. 

At the same time, considerable development is necessary in order to efficiently access resources at HPCs. 
Most workflows directed to HPCs have required limited I/O and external connectivity, (e.g. Monte Carlo simulation) and produced a modest amount of output.
To enable more workflows to access HPC computing facilities, particularly those which have very restricted network input and output, research into the optimized facility configuration should be performed to provide the needed capabilities for HEP workflows. All of these capabilities currently exist in some form, but need to be improved in terms of scalability and accessibility. We envision a focus on development efforts on the following components and capabilities to help users effectively utilize HPC facilities: Code distribution, database access, data ingress and egress from HPC facilities, monitoring, campaign- and data-aware Workload Management, and development access to HPC. The integration of these components into justIN, glideInWMS, and HTCondor are envisioned as primary features in future developments. By leveraging GlideInWMS and HEPCloud, justIN has access to DOE’s NERSC GPU cluster and enables flexible and automated job allocation, dynamically provisioning computing resources based on real-time demand. This is particularly valuable for AI/ML training and inference tasks described in the next section, where computational workloads can vary significantly. In the FNAL Glidein factory list, the EU and UK's UK\_QMUL and ES\_PIC GPU farms are already capable of supporting DUNE jobs. In summary, DUNE is already capable of utilizing GPU resources in distributed computing for AI/ML training and evaluation, but development is necessary to ensure the efficiency and seamless utilization of those resources.

\section{DUNE Software Research and Development}

\subsection{Community Reconstruction Algorithm Development}
Currently, DUNE reconstruction and simulation processing has built heavily upon the algorithms developed both within the liquid argon TPC community (e.g. ArgoNeuT, LArIAT, MicroBooNE, ICARUS, SBND, etc.), and from the integration of reconstruction software suites originating from other communities. In particular, the integration of Pandora and WireCell, within the LArSoft ecosystem, has been an essential part of reconstruction and simulation tools for DUNE. These algorithms have enabled robust sensitivity projections for DUNE, physics measurements in ProtoDUNE, and detector optimization studies.

The UK-led Pandora Software Development Kit [1] and algorithm libraries perform reconstruction of neutrino interactions in Liquid Argon Time Projection Chamber (LArTPC) detectors [2, 3]. European collaborators have implemented Pandora in DUNE, where it serves as the primary event reconstruction framework. Pandora plays a crucial role in identifying and classifying neutrino interactions, leveraging a multi-algorithm approach to efficiently disambiguate detector signals, reconstruct particle trajectories, and identify interaction vertices. Key components of the Pandora reconstruction framework include cosmic-ray background rejection, track and shower classification, vertex reconstruction, particle identification (PID), and clustering techniques. The continued robust support of a broad community of reconstruction algorithms within Europe is important to taking full advantage of the high-precision data produced by the DUNE near and far detectors.

\subsection{Integration of AI/ML within DUNE Software}

Artificial Intelligence and Machine (Deep) Learning (AI/ML) is prevalent throughout many aspects of DUNE. A wide range of techniques have been employed in many diverse use cases, many of which have been developed by collaborators in Europe. 

AI/ML is applied in key reconstruction steps, including signal processing, hit finding, clustering, kinematics, event classification, particle identification, and non-beam neutrino reconstruction. DUNE’s LArTPC is at the forefront of AI/ML development for reconstruction due to its complexity and high resolution, where traditional analytic methods can face challenges. Beyond reconstruction, AI/ML is widely utilized in simulation, data acquisition (DAQ), data processing, beam design and monitoring, as well as quality assurance and control (QA/QC). Additionally, DUNE has contributed to enhancing AI infrastructure and aligns with priority research directions for AI/ML in high-energy physics. European groups have played a pivotal role in advancing AI/ML development in DUNE, contributing significantly to key methodologies and applications across the experiment. A number of examples are discussed below.

The previously mentioned Pandora Software Develop Kit utilizes various AI/ML techniques to boost its performance in key reconstruction steps. Examples include the use of Boosted Decision Trees (BDTs) for labelling slices as cosmics or neutrino/beam events, as well as for track and shower identification. Convolutional Neural Networks (CNNs) and BDTs are applied to vertexing tasks [4], while CNNs are also used for particle identification (PID) in Pandora. Additionally, semantic segmentation and Graph Neural Networks (GNN) are being explored for track and shower clustering, with potential applications in vertexing [5]. Building on its success as DUNE’s primary event reconstruction framework, future developments in Pandora will focus on enhancing AI/ML integration, optimizing reconstruction performance, and expanding applicability to new physics analyses.

European groups have played a leading role in developing the Convolutional Visual Network (CVN), a key machine learning algorithm used in the DUNE experiment for neutrino event classification [6]. The DUNE CVN processes images of neutrino interactions in the LArTPC to predict neutrino flavor, achieving classification efficiencies of 90\% for electron neutrinos and 95\% for muon neutrinos. This European-led innovation significantly enhances event selection purity and reconstruction accuracy, outperforming traditional techniques. Furthermore, the CVN classifier has been instrumental in demonstrating that DUNE can achieve the CP-violation sensitivity outlined in its Conceptual Design Report (CDR). Since the CVN significantly improved the identification and differentiation of neutrino and antineutrino interactions based on their reconstructed energy distributions, it has been crucial in maximizing DUNE's oscillation analysis capabilities and advancing its major physics goal of probing CP violation in the neutrino sector.
A CNN was developed to identify hits coming from tracks and showers in ProtoDUNE [7] using a small patch approach as opposed to a U-Net for semantic segmentation. It was the first Deep Learning algorithm developed within the DUNE Collaboration to be tested on data, and demonstrated that similar performance can be obtained when classifying hits in simulation and data. This was an important milestone as it showed that it is possible to avoid potential domain-shift biases in these algorithms and improved trust in Deep Learning approaches generally.

\section{Neutrino-Nucleon Interaction Simulations}

The broad physics program of DUNE is supported by a range of neutrino interaction Monte Carlo generators with the primary being GENIE\cite{Andreopoulos:2009rq,GENIE:2021npt} but others like NuWro\cite{Nuwro,Nowak:2006sx}, GiBUU\cite{Buss:2011mx, Mosel} and NEUT\cite{Mosel} are also utilized directly in DUNE simulation chain or via the NUISANCE \cite{Stowell:2016jfr} comparison framework. 
Each of the neutrino generators offer different approaches to various aspects of the interactions. As DUNE will use a wideband beam, the neutrino interaction generators need to properly simulate all the possible channels spanning about an order of magnitude of neutrino energy (1-10GeV), avoiding double counting, optimizing for rare processes and combining nuclear effects and Final State Interactions. There is no sufficient neutrino interactions  data to disambiguate some of the effects, e.g. nucleon and nuclear structure, and the generators are tested with electron scattering data. 
DUNE will require multiple generators to deliver its ambitious physics program to detect biases and correctly account for systematic uncertainties. 
 
Each of the generators has been developed over the last 20–30 years and are supported by a small group of developers. GENIE has support from Fermilab, as a primary generator for many of Fermilab-hosted experiments. But other generators are related to DUNE via membership in the collaborations.
One of the challenge for the development of generators is a lack of long-term support for the primary developers and their leaders, as it is not well aligned with the typical academic career path. 

GENIE, NuWro and GiBUU are generators that were primarily developed in European institutions. The availability and utilization of multiple generators enables DUNE to fully develop the physics program, optimize the beam and detector configurations, with minimal bias from any one generator or interaction model. Continued support for numerous generators and the developers is seen as critical to fully realize DUNE's physics potential.

\section{Conclusion}

European contributions to DUNE computing have and will be essential to the success of the DUNE physics program. These European contributions of computing resources, strong operational support, infrastructure for distributed computing, and algorithmic development have allowed DUNE to face the current computing challenges ahead of data taking, and plan to successfully address the challenges faced by the full experiment. The continued involvement of European institutions, researchers, and computer scientists in DUNE will be critical for the success of the experiment and answering some of the most challenging questions in neutrino and particle physics.

\end{document}